\documentclass{aastex63}
\usepackage{CJK}
\shorttitle{Compressible Turbulence}
\shortauthors{Fu et al.}

\begin{document}
\begin{CJK*}{UTF8}{gbsn}

\title{Nature and Scalings of Density Fluctuations of Compressible MHD Turbulence 
with Applications to the Solar Wind}

\correspondingauthor{Xiangrong Fu}
\email{sfu@newmexicoconsortium.org}

\author[0000-0002-4305-6624]{Xiangorng Fu}
\affiliation{New Mexico Consortium \\
Los Alamos, NM 87544, USA}

\author[0000-0003-3556-6568]{Hui Li (李晖)}
\affiliation{Los Alamos National Laboratory \\
Los Alamos, NM 87545, USA}

\author{Zhaoming Gan}
\affiliation{New Mexico Consortium \\
Los Alamos, NM 87544, USA}

\author{Senbei Du}
\affiliation{Los Alamos National Laboratory \\
Los Alamos, NM 87545, USA}

\author{John Steinberg}
\affiliation{Los Alamos National Laboratory \\
Los Alamos, NM 87545, USA}



\begin{abstract}
The solar wind is a magnetized and turbulent plasma. Its turbulence is 
often dominated by Alfv\'enic fluctuations and often deemed as nearly incompressible 
far away from the Sun, as shown by in-situ measurements near 1AU. 
However, for solar wind closer to the Sun, the plasma $\beta$ decreases 
(often lower than unity) while the turbulent Mach number $M_t$ increases
(can approach unity, e.g., transonic fluctuations). 
These conditions could produce significantly more compressible effects, 
characterized by enhanced density fluctuations, as seen by several space missions. 
In this paper, a series of 3D MHD simulations of turbulence are carried out to 
understand the properties of compressible turbulence, particularly the generation of 
density fluctuations. We find that, over a broad range of parameter space
in plasma $\beta$, cross helicity and polytropic index, the turbulent density fluctuations
scale linearly as a function of $M_t$, with the scaling coefficients showing
weak dependence on parameters. Furthermore, 
through detailed spatio-temporal analysis, we show that
the density fluctuations are dominated by low-frequency nonlinear structures, 
rather than compressible MHD eigen-waves. 
These results could be important for understanding how compressible turbulence
contributes to solar wind heating near the Sun. 

\end{abstract}


\section{Introduction} \label{sec:intro}

The solar wind is a high-speed plasma flow emitted from the solar surface, 
as first predicted by \citet{parker_apj_1958}. It was soon discovered that 
the solar wind is turbulent, containing fluctuations in a vast range of spatial and temporal scales.
The free energy source for solar wind turbulence is typically at large scales, 
associated with activities near the Sun. (Energy can also be injected by instabilities 
at kinetic scales driven by beams or temperature anisotropies.) These large-scale 
perturbations will interact nonlinearly, causing cascade of energy to smaller 
scales and forming power-law spectra of magnetic, velocity and density fluctuations, 
which are believed to contribute to the solar wind heating. 

In the classic picture of solar wind turbulence \citep[see a recent review by][]{bruno_lrsp_2013}, 
the fluctuations are treated as Alfv\'enic and incompressible, i.e., the divergence 
of velocity fluctuations is zero. 
Theories have been developed to deal with nearly incompressible turbulence  
in the solar wind and they have explained many
of the key features of solar wind turbulence \citep{montg_jgr_1987, matth_pof_1988, matth_jgr_1991,
zank_jgr_1992, zank_apj_2017}. 
However, there exists a non-negligible component of compressible turbulence 
in the solar wind throughout the heliosphere, characterized by enhanced density 
fluctuations. Furthermore, there have been several theoretical and numerical studies 
on compressible MHD turbulence in astrophysical plasmas as well. 
Radio wave scintillation observations reveal a nearly Kolmogorov
spectrum of density fluctuations in the ionized interstellar medium, 
suggestive of compressible MHD turbulence.  Past theory and simulations 
have suggested that 
the density fluctuations are due to the slow mode and the entropy mode 
\citep{lithw_apj_2001}. Many numerical simulations have been used 
to understand the nature of compressible MHD turbulence as well
\citep{cho_mnras_2003, kowal_apj_2007, yangy_pof_2017, Shoda2019, makwa_prx_2020}.
In addition, some previous studies have revealed that, 
due to different properties of MHD waves, compressible perturbations 
(associated with fast and slow waves) cascade differently than 
incompressible perturbations (associated with Alfv\'en waves). 
For example, MHD simulations by \citet{cho_prl_2002, cho_mnras_2003} 
show fast-mode turbulence tends to form an isotropic spectrum (though 
fast-mode turbulence seems to be anisotropic in the kinetic regime, as shown by 
\citet{svidz_pop_2009}), while Alfv\'enic turbulence tends to form an 
anisotropic spectrum with respect to the background magnetic field.

It is worth noting that nonlinear interactions among 
incompressible and compressible components make the study of compressible 
turbulence very challenging. 
In general, the density fluctuations in compressible MHD turbulence can originate from different processes. In the limit of regarding fluctuations as MHD waves, the variations
are expected to follow the $\omega-{\bf k}$ dispersion relation of magnetosonic fast and slow modes, 
and fluctuating density, velocity and magnetic fields satisfy relative relation set by 
eigenvectors in linear theory. In fact, three MHD eigen-modes (fast, slow and Alfv\'en modes) can 
couple with each other nonlinearly through wave-wave 
interactions \citep[e.g.,][]{chand_prl_2005}. 
These processes typically
render an MHD system with a finite $\beta$ compressible
(where $\beta$ is the ratio of plasma pressure over the magnetic pressure), even if
it starts with incompressible velocity fluctuations. 
Another way to produce density variations is via instabilities. 
One example of such processes
is the parametric decay instability, where a large amplitude Alfv\'en wave 
decays into another Alfv\'en wave and a slow magnetosonic wave 
\citep{derby_apj_1978,golds_apj_1978}, which is a source of density 
fluctuations . This process is of particular
relevance to the solar wind near the Sun  
\citep{2015JPlPh..81a3202D, shi_parametric_2017, Reville2018, 
Shoda2018, fu_apj_2018, Tenerani2020}.

Density fluctuations could, of course, simply arise from the
turbulence driving process, then they will continue cascade to different
scales from nonlinear interactions. This process is 
not necessarily associated with MHD waves. In fact, whether and how much
MHD waves exist in a turbulence is a non-trivial issue. 
One approach is to examine whether fluctuations follow the 
dispersion relation of MHD waves (which is very challenging in general). 
Here, we use expressions such as waves and nonlinear structures to 
describe whether fluctuations will follow the eigen-mode dispersion relations or not. 
An example of known nonlinear structures is the pressure balance structure (PBS)
\citep{tu_ssr_1995,bruno_lrsp_2013} in the solar wind, 
where the sum of the magnetic and thermal 
pressures is nearly constant across the structure. PBS is non-propagating 
(though it can be convected by the solar wind) and static, having a zero 
frequency in the plasma frame. PBS can contain spatial variation of density, 
which will be detected as a time series of density fluctuation when  
spacecraft is traveling through the the solar wind plasma. 
In principle, the fluctuations in compressible MHD turbulence do not need
to satisfy the full dispersion relationships of eigen-modes. Indeed, 
recent studies of compressible turbulence by \citet{gan_ApJ_2022} 
showed that velocity and magnetic fluctuations were dominated by nonlinear structures 
rather than MHD waves in the simulations using an advanced spatio-temporal 
4D FFT analysis. This conclusion is quite different from many previous analyses where
the decomposition of fluctuations is based solely on the spatial variations 
without their temporal properties \citep[see also][]{dmitr_pop_2009, andres_pop_2017,yang_mnras_2019,brodi_apj_2021}.
In this paper, we will also use this 4D FFT approach to examine whether 
the density fluctuations are associated with MHD waves or nonlinear structures.

One key parameter that controls the compressibility of the turbulence is 
the turbulent Mach number $M_t$, which is the ratio of total velocity fluctuation $\delta v$ 
to the typical sound speed 
$c_s=\sqrt{\gamma p/\rho}$ (where $\gamma$ is the polytropic index). 
Since $M_t=\delta v/c_s=(\delta v/v_A) \sqrt{2/(\beta \gamma)}$, 
where $\beta = p/(B^2/8\pi)$, 
compressible turbulence is more easily generated in  low $\beta$ plasmas, 
given a similar level of fluctuation $\delta v/v_A$ (or $\delta B/B_0$ for Alfv\'enic perturbation). 
This makes the solar wind close to the Sun especially interesting because $\beta$ 
is expected to be lower than unity, compared to about unity at 1 AU \citep{gary_sp_2001}. 
The Parker Solar Probe (PSP) mission is exploring the near Sun region \citep{fox_ssr_2016}, 
enabling new discovery and understanding of turbulence in the solar wind 
\citep{chen_apjs_2020,chen_a&a_2021, shi_A&A_2021}. Recently, 
PSP has entered the sub-Alfv\'enic region and start to observe turbulence 
in low-$\beta$ solar wind \citep{kasper_prl_2021}. 

There have seen several studies on the relation between density fluctuations $\delta \rho$ 
and the turbulent Mach number $M_t$. Nearly-incompressible MHD (NI-MHD)
theory \citep{matth_pof_1988, matth_jgr_1990,zank_pof_1993} 
predicts $\delta \rho \propto M_t^2$ if 
$M_t$ is a small parameter, similar to nearly-incompressible theory in 
neutral fluids. Numerical simulations  of fluid turbulence have confirmed 
that the $\delta\rho \propto M_t^2$ scaling holds at small $M_t$ and 
it transitions to more complex scaling at large $M_t$ \citep[e.g.][]{cerre_pof_2019}.
There was observational evidence in the solar wind that supports 
$M_t^2$ scaling and NI-MHD theory \citep{matth_jgr_1991}, but other studies 
found a lack of clear scaling \citep{tu_jgr_1994,bavas_jgr_1995}, 
possibly due to the existence of inhomogeneous structures \citep{bruno_lrsp_2013}. 
Given that PSP is now providing the in-situ measurements of compressible 
turbulence, we have performed MHD turbulence simulations, 
particularly with relatively high $M_t$,  to investigate how 
various turbulence quantities correlate among themselves.

The paper is organized as follows. Sec.\ref{sec:sim} describes the numerical 
model used in this study. Simulations results and analyses are presented in detail in Sec.\ref{sec:res}. 
We summarize main results in Sec.\ref{sec:dis}, with discussion of unresolved 
issues and possible future directions.

\section{Numerical Simulation} \label{sec:sim}

To study compressible turbulence and density fluctuations in the context of the 
solar wind in the inner heliosphere ($<1$ AU), we carry out a series 3D compressible MHD simulations.
We select plasma parameters that represent typical solar wind conditions within 1 AU. 
We use the high-performance 
code ATHENA++ \citep{stone_apjs_2020} to solve the ideal compressible MHD equations:
\begin{equation}
\label{eq:mass}
\frac{\partial\rho}{\partial t} + \nabla\cdot(\rho{\bf v}) = 0,
\end{equation}
\begin{equation}
\label{eq:momentum}
\frac{\partial (\rho {\bf v})}{\partial t} + \nabla\cdot \left( \rho {\bf v v} + p{\bf I} + B^2{\bf I}/2 - {\bf B B}\right)  = {\bf f}_v,
\end{equation}
\begin{equation}
\label{eq:induction}
\frac{\partial {\bf B}}{\partial t} - \nabla\times( {\bf v} \times {\bf B}) = {\bf f}_b.
\end{equation}
\begin{equation}
\label{eq:energy}
\frac{\partial E}{\partial t} + \nabla\cdot\left[\left(E +p + B^2/2 \right){\bf v} -{\bf B}({\bf v}\cdot{\bf B})\right] = 0,
\end{equation}

\noindent
with the polytropic equation of state $p \rho^{-\gamma}=\rm const$, 
where $\gamma$ is the polytropic index. Since we are interested in local properties 
of turbulence driven by large-scale free-energy input, our model does not include 
global structure of the solar wind. Instead, we assume a system with 
uniform background magnetic
field and periodic boundary conditions. The simulation domain 
is an elongated box with $L_x=8\pi$, $L_y=L_z=2\pi$ and the background 
magnetic field is $B_0 {\bf e}_x$. Periodic boundary conditions are applied in 
all three directions. The Alfv\'en crossing time is $\tau_A=L_x/v_A=8\pi$. 
For most of our runs we use 512 cells to resolve each dimension. 

Starting from a quiet uniform plasma ($\rho_0=1$, $B_{0}=1$ , and zero fluctuations), 
we perturb the MHD system  by adding driving terms in both momentum and magnetic field 
equations (Eq. \ref{eq:momentum} and Eq. \ref{eq:induction}).
We maintain the divergence free condition for the magnetic field by 
keeping $\nabla \cdot {\bf f}_b=0$ to machine precision \citep{gan_ApJ_2022}. 
For all of the runs, we also keep $\nabla \cdot {\bf f}_v=0$ to model incompressible 
driving at large scales. ${\bf f}_v$ and ${\bf f}_b$ are applied at large scales 
where their wave number $k\leq k_{\rm inj} = 3$.
By varying the relative strength and phase of ${\bf f}_v$ and ${\bf f}_b$, 
we are able to vary the Alfv\'en ratio $r_A\equiv|\delta {\bf v}|^2/|\delta {\bf B}|^2$ 
and cross helicity $\sigma_c \equiv(|{\bf z}^+|^2-|{\bf z}^-|^2)/(|{\bf z}^+|^2+|{\bf z}^-|^2)$ (where ${\bf z}^{\pm}\equiv \delta {\bf v}\pm \delta {\bf B}$ are Els\"{a}sser variables)
of the injected fluctuations. To some degree, they replicate the observed solar wind conditions.
By keeping the relative phase and strength of ${\bf f}_v$ and ${\bf f}_b$ and changing the 
magnitude of both quantities proportionally, we can also vary the driving 
strength and achieve different levels of turbulence.
Typically, an adiabatic equation of state is adopted with $\gamma=5/3$ to simulate 
MHD turbulence in the solar wind. But we will vary $\gamma$ and study its effect on 
compressible turbulence and density fluctuations, as inspired by recent observations 
from PSP \citep{nicol_apj_2020}. $\gamma> 5/3$ implies heating of the system 
by an external energy source (not included in the model) and $\gamma< 5/3$ implies cooling. 

We should emphasize that, even though we have tried to keep the driving 
process $\nabla \cdot {\bf f}_v=0$, this does not mean that $\nabla \cdot ({\rho {\bf v}}) \approx 0$
during the evolution.
For example, because we have used ${\bf f}_b$, the ``added" Lorentz force 
${\bf J}\times {\bf B}$ due to injection is typically non-zero. This leads to momentum
variations that are not guaranteed to be $\nabla \cdot ({\rho {\bf v}}) =0$. Although
$\nabla \cdot {\bf f}_v=0$ is satisfied through the injection process, the
density variations can be excited through the joint injection process of  ${\bf f}_v$ and ${\bf f}_b$.
We will discuss this point in more details in the next section. 

\begin{deluxetable}{c|ccc|ccccc}
  \tablecaption{Main input parameters (column 2-4) and turbulent quantities (column 5-9) for 3D MHD simulations. The system size is $8\pi \times 2\pi \times 2\pi$ with a uniform magnetic field $B_0=1$ in the $x$ direction. 
Turbulent quantities are measured in the late stage of each simulation when the system is in a quasi-steady state.
  \label{tab:para}}
  \tablecolumns{8}
  \tablehead{
  \colhead{Run} &
  \colhead{number of cells} &
  \colhead{$\beta_0$} &
  \colhead{$\gamma$} &
  \colhead{$M_A$} &
  \colhead{$M_t$} &
  \colhead{$\delta \rho$ }&
  \colhead{$\sigma_c$ } &
  \colhead{$\beta$ }
  }
  \startdata
     1 & $512^3$ & 1.0 & 1.67 &0.25 & 0.27 & 0.13 & 0.93 & 1.03 \\
    2  & $512^3$ & 1.0 & 1.67  &0.34 & 0.37 & 0.19 & 0.85 & 1.07\\
    3  & $512^3$ & 1.0 & 1.67  &0.17 & 0.19 & 0.09 & 0.91 & 1.01 \\
     4& $512^3$ & 0.2 & 1.67  &0.24 & 0.59 & 0.33 & 0.67 &0.24 \\
     5 & $512^3$ & 1.0 & 1.67  &0.21 & 0.23 & 0.11 & -0.20 & 1.03 \\
     6  & $512^3$ & 1.0 & 2.7  &0.28 & 0.24 & 0.10 & 0.93 & 1.08 \\
     7 & $1024^3$ & 1.0 & 1.67  &0.28 & 0.30 & 0.13 & 0.92 & 1.03 \\
8  & $256^3$ & 1.0 & 1.67  &0.24 & 0.26 & 0.12 & 0.92 & 1.03
  \enddata
\end{deluxetable}

\section{Results} \label{sec:res}

The key parameters and turbulent quantities of our MHD
simulations are summarized in Table \ref{tab:para}. Note that, in addition to these listed runs,
we have also performed a few runs with higher  resolution ($1024^3$)
and many lower resolution ($256^3$) runs to map out the parameter space (see below), 
as well as to check the consistency of the results.
In this section, we will first describe the evolution 
of turbulent quantities under different driving conditions. Second, 
we present the scaling of density fluctuations as a function of turbulent Mach number 
$M_t$ and examine how the scaling varies with key plasma/turbulence parameters. 
Third, we investigate the nature of the density fluctuations and their relation to 
compressible MHD waves
and nonlinear structures.

\subsection{Evolution of Turbulent Quantities}

\begin{figure}
    \centering
    \includegraphics[width=0.48\textwidth]{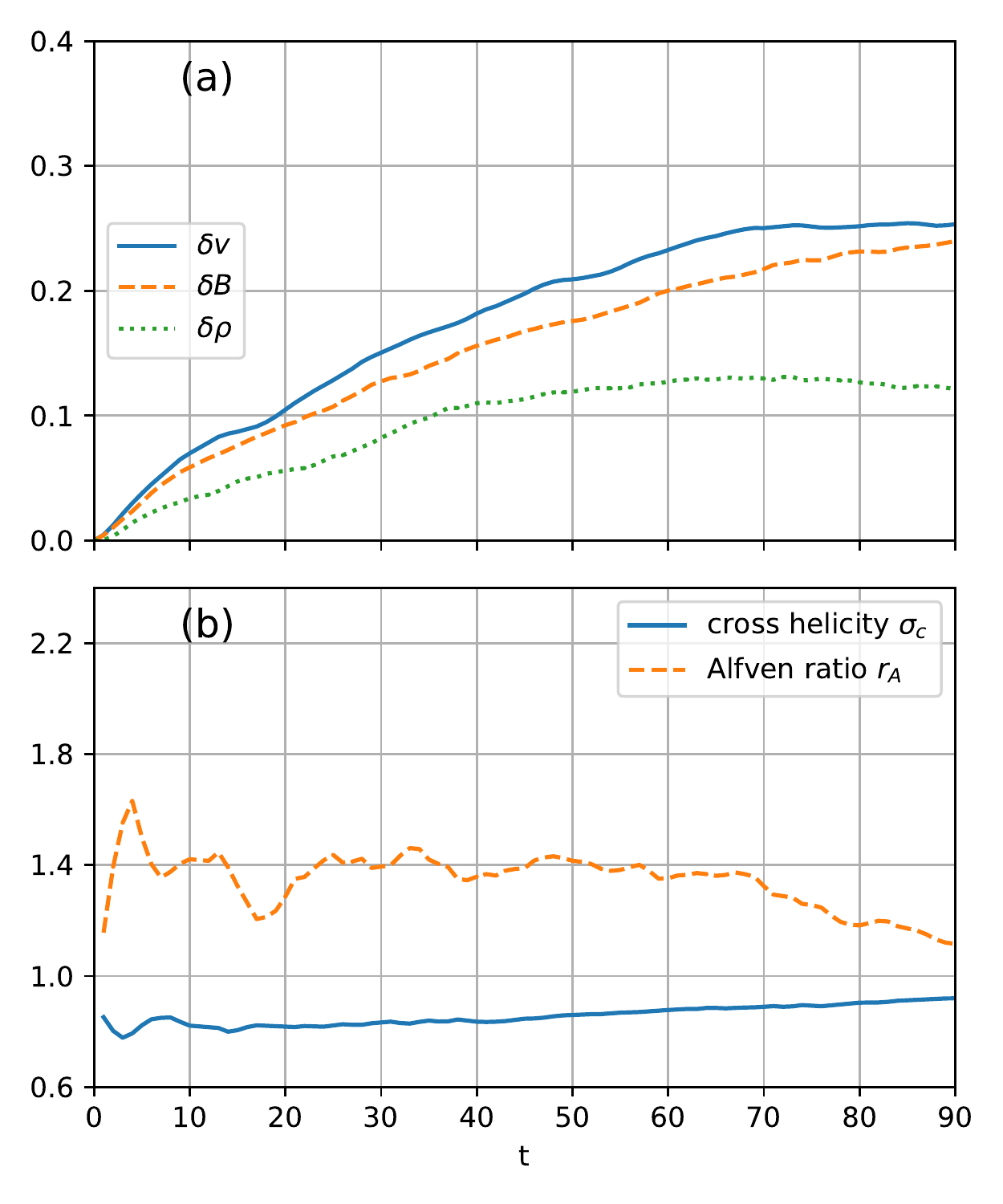}
    \includegraphics[width=0.48\textwidth]{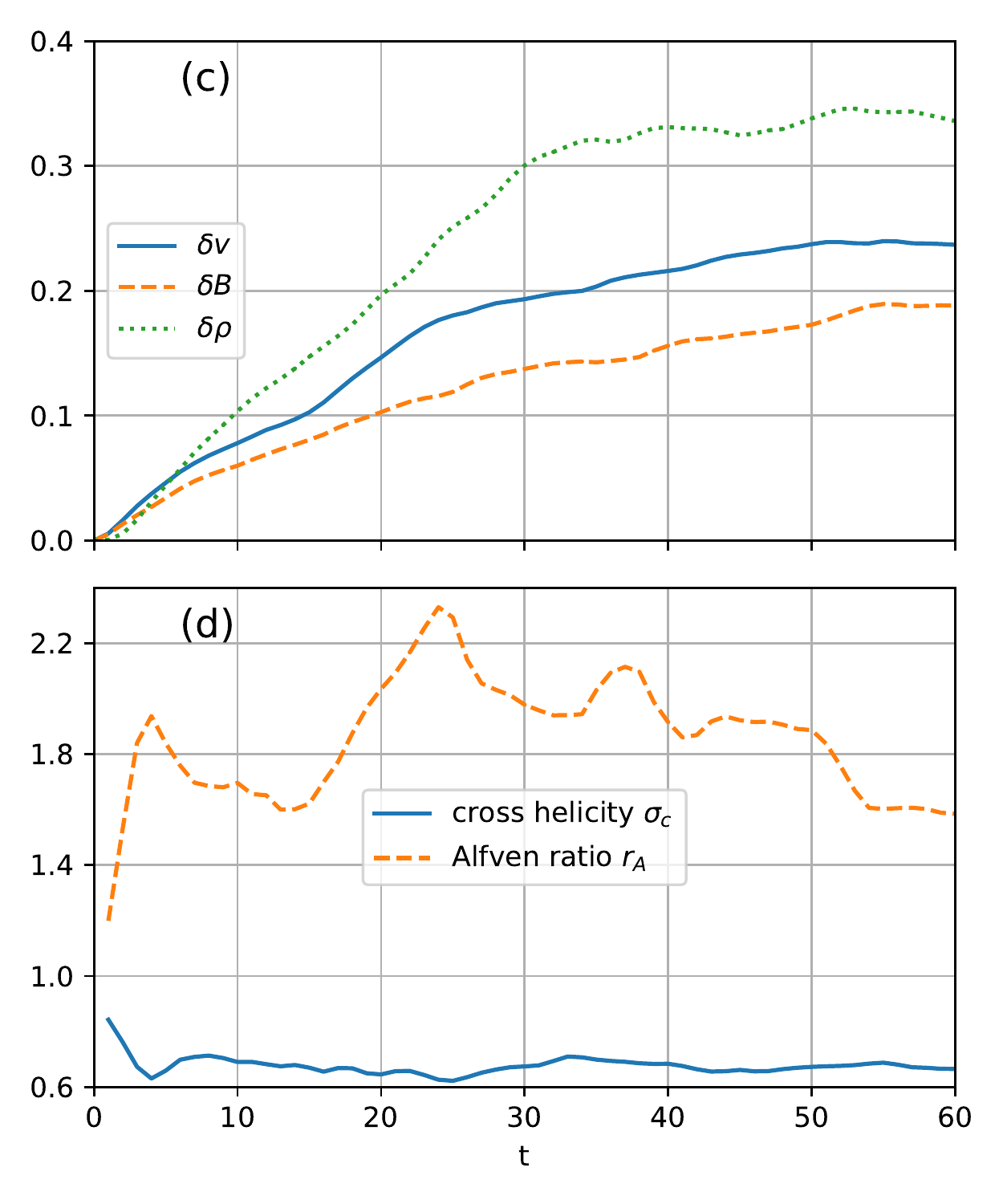}
    \caption{Run 1: $\beta_0=1.0$, (a) time evolution of magnetic field, velocity field and density fluctuations. The fluctuation level saturates after about $t=50$ when the system enters into a quasi-steady state;  (b) time evolution of cross-helicity $\sigma_c$, and Alfv\'en ratio $r_A$. Run 4: $\beta=0.2$, (c-d) Similar to panels (a-b) but for Run 4 with $\beta_0=0.2$. }
    \label{fig:fluc}
\end{figure}

Our nominal case is Run 1 with an initial $\beta_0 =1 $ and an adiabatic 
equation of state (i.e., $\gamma=5/3$). 
The system is driven by Alfv\'enic incompressible perturbations  
at large scales (with $k_{\rm inj}=3$) and it has a final cross helicity  $\sigma_c=0.93$. 
Figure \ref{fig:fluc} (a) shows the time history of averaged 
velocity field, magnetic field and density fluctuations in Run 1, with
$$\delta B\equiv\langle |{\bf B}-{\bf B}_0|\rangle/B_0,\,
\delta v\equiv \langle |{\bf v}-{\bf v}_0|\rangle/v_A,\,
\delta \rho\equiv \langle \rho -\rho_0\rangle/\rho_0$$
where the brackets indicate the root-mean-square value over the whole simulation domain. 
These quantities are measures of the overall level of turbulence. 
With a continuous energy injection, all three fluctuation quantities increase 
gradually as a function of time. After an initial growth stage ($0<t<60$), 
the turbulence reaches a quasi-steady state with a saturated level of fluctuations. 
Between $t=70$ and $t=90$, we check the spectra of $\delta v$, $\delta \rho$ and $\delta B$ 
and confirm that they remain nearly unchanged. 
Since we use an adiabatic equation of state, the total energy of the system 
is increasing with the continuous injection. After the turbulence is saturated, 
most of the input energy is converted into the internal energy of plasma, 
causing plasma heating in the later stage of the simulation ($\beta$ increases 
from 1.0 at $t=0$ to $\sim 1.03$ at $t=90$). 
Overall, the turbulence and plasma parameters are roughly 
controlled by the injection.

For Run 1, $M_t$ reaches 0.27 in the quasi-steady state and $\delta \rho$ reaches 0.13. 
By varying the strength of driving forces (${\bf f}_b$ and ${\bf f}_v$), 
we can obtain turbulence with different levels of $M_t$. 
In Run 2 (3), which uses the same parameters as Run 1 except stronger (weaker) driving 
forces, we obtain a larger (smaller) $M_t=0.37$ ($M_t=0.19$) and stronger (weaker) density 
fluctuations $\delta \rho=0.19$ ($\delta \rho=0.09$) (see Table \ref{tab:para} for details).

Note that although we keep the injected fluctuations Alfv\'enic with
$\delta v=\delta B$, the plasma has freedom to respond to the 
injection and total $\delta v$ is slightly larger than $\delta B$ throughout the simulation. 
This can also be seen in Figure \ref{fig:fluc} (b), where $r_A$ first increases from $1.2$
to $1.6$ then reduces to $1.1$ ($r_A=1$ for Alfv\'enic fluctuations) at the later stage. 
Similarly, we intend to have an imbalanced turbulence with $\sigma_c=0.9$, 
the resulting  $\sigma_c$ varies from $\sim 0.8$ at the beginning to $\sim 0.9$ 
at the end of the simulation.

To contrast with Run 1, we carry out Run 4 with the same parameters 
except a lower $\beta_0 =0.2$.
This represents the solar wind condition much closer to the Sun. 
As shown in Figure \ref{fig:fluc} (c), the fluctuations saturate at 
$t\sim 40$, with a similar level of velocity fluctuation ($\delta v\sim 0.24$) 
to Run 1. 
It is notable that the density fluctuation generated in Run 4 is much 
higher than Run 1, 
with $\delta \rho\sim 0.33$ (vs. $0.13$ in Run 1). 
This is the characteristic response of low-$\beta$ plasma. 
Because we hold $M_A = \delta v/v_A$ the same for both runs, 
Run 4, being lower $\beta$, achieves much higher 
turbulence March number $M_t$ with same $\delta v$.
Another feature of Run 4, different from Run 1, is that 
the Alfv\'en ratio remains higher than unity ($\sim 1.5$ at the later stage, 
i.e. more energy goes into the velocity fluctuation than the magnetic fluctuation) 
throughout the simulation, even though the driving forces are Alfv\'enic. 
Similarly, the cross helicity remains around 0.7, lower than the value of Run 1. 
These quantities can be used to study the 
possible relationship between  Alfv\'enicity and  compressibility 
in the solar wind too \citep{bruno_lrsp_2013}.

\begin{figure}
    \centering
    \includegraphics[width=0.9\textwidth]{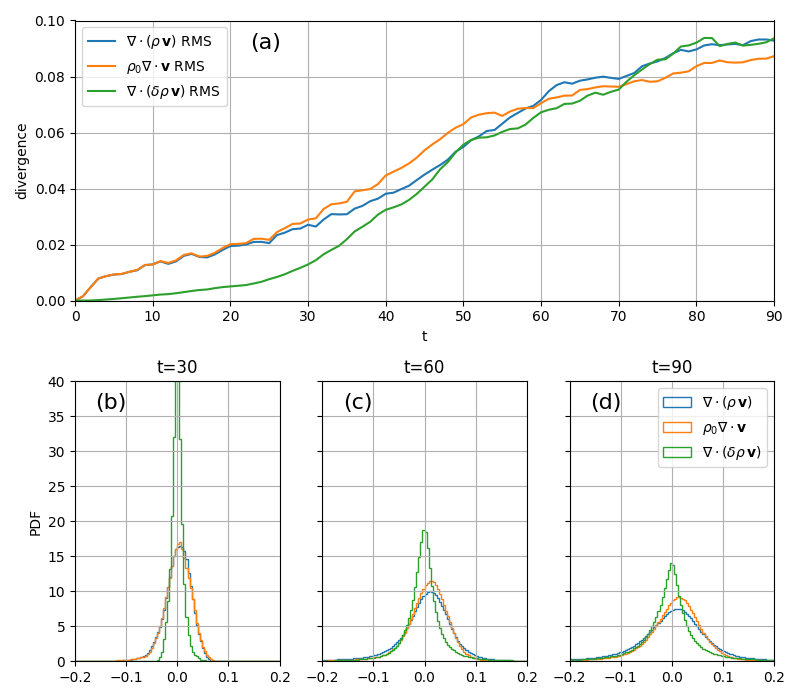}
    \caption{Run 1: (a) History of the root-mean-square values of the divergence terms that contribute to growth of density fluctuations, and histograms of the divergence term PDFs at (b) t=30, (c) t=60 and (d) t=90. In the growing phase of turbulence ($t<60$), the total divergence $\nabla\cdot (\rho {\bf v})$ is dominated by the linear term $\rho_0 \nabla \cdot {\bf v}$. In the saturation phase, the nonlinear term $\nabla \cdot (\delta \rho\, {\bf v})$ is also quite strong. }
    \label{fig:fluc2}
\end{figure}

As we emphasized earlier, even though the driving
ensures that $\nabla \cdot {\bf f}_v=0$ and $\nabla \cdot {\bf f}_b=0$, the evolution
of both magnetic field (thus Lorentz force) and momentum does not
guarantee that $\nabla \cdot (\rho {\bf v}) =0$. 
Using Run 1 as an example, Figure \ref{fig:fluc2} depicts the 
evolution of RMS values of various
terms in the continuity equation (panel a) along with their probability density functions (PDFs) in panels b-d. 
(These quantities are calculated using the cell-based values.)
It can be seen that, even though $\nabla \cdot {\bf v} \approx 0$ when integrated over volume (i.e., the averaged value is near zero),
there are large variances for  $\nabla \cdot {\bf v}$ and $\nabla \cdot (\delta \rho {\bf v})$,
both of which contribute to large $\nabla \cdot (\rho {\bf v})$ variations. 
It is also interesting to notice that the $\nabla \cdot {\bf v}$ term has major contributions
throughout the evolution (being dominant in the beginning), the 
$\nabla \cdot (\delta \rho\, {\bf v})$ term also becomes important as time goes on,
even becoming the bigger term after $t > 70$. 

\subsection{Scaling of Density Fluctuations}

As shown above, the density variations depend strongly on the turbulent Mach number $M_t$
and show some dependence on plasma $\beta$ and other parameters. 
To systematically study density fluctuations under various conditions relevant to the solar wind, 
we explore  a large parameter space with different $\beta$, $\sigma_c$, $\gamma$ and $M_t$.  
Before we present our main results, let us first examine the spectral properties 
of density fluctuations in the nominal case. 

\begin{figure}
    \centering
    \includegraphics[width=0.9\textwidth]{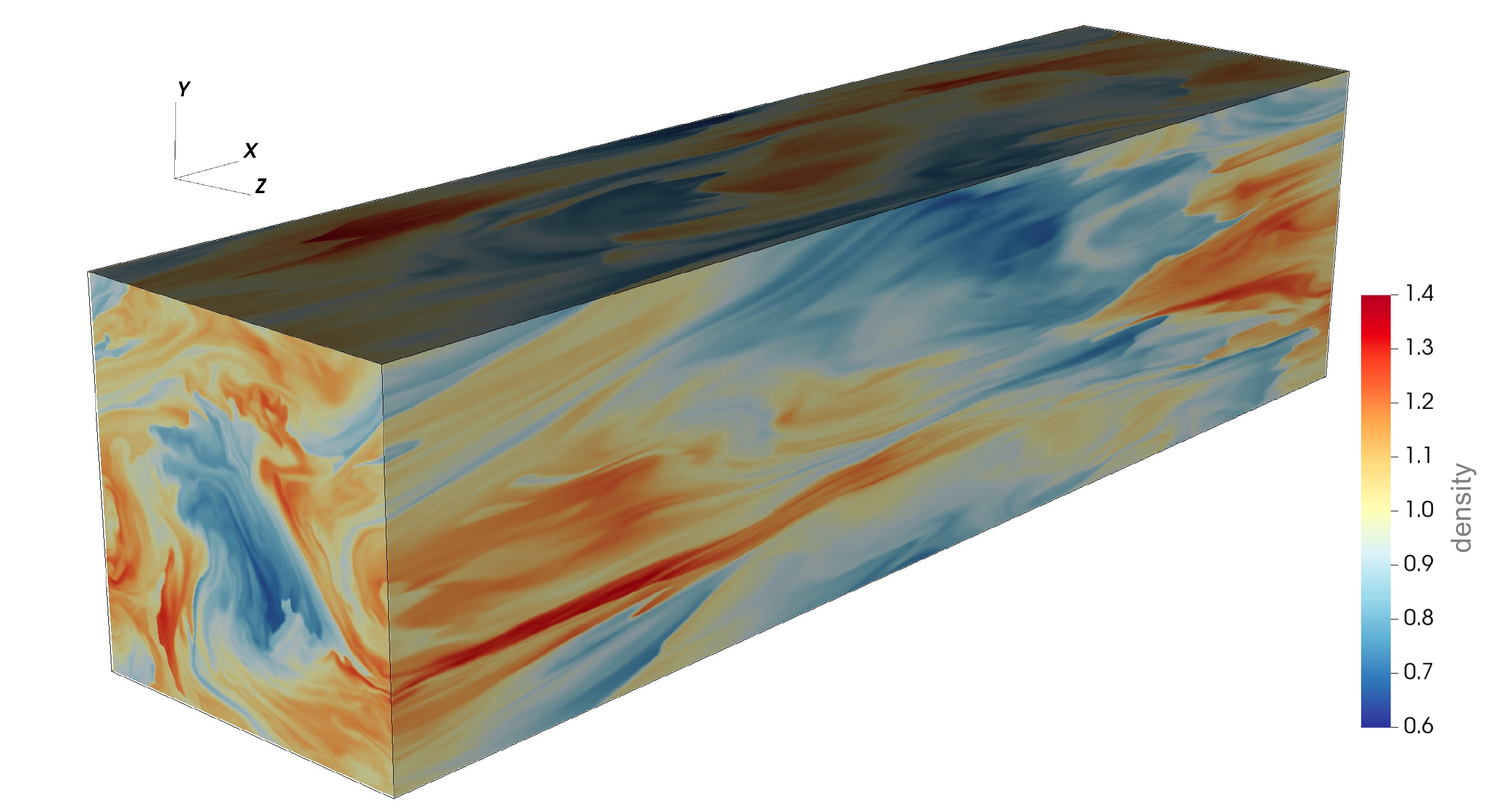}
    \caption{ 3D snapshot of density fluctuation at $t=80$ for Run 1. The density fluctuations are elongated along the direction of background magnetic field ($x$), showing strong anisotropy.
    }
\label{fig:3dsnap}
\end{figure}

\begin{figure}
    \centering
     \includegraphics[width=\textwidth]{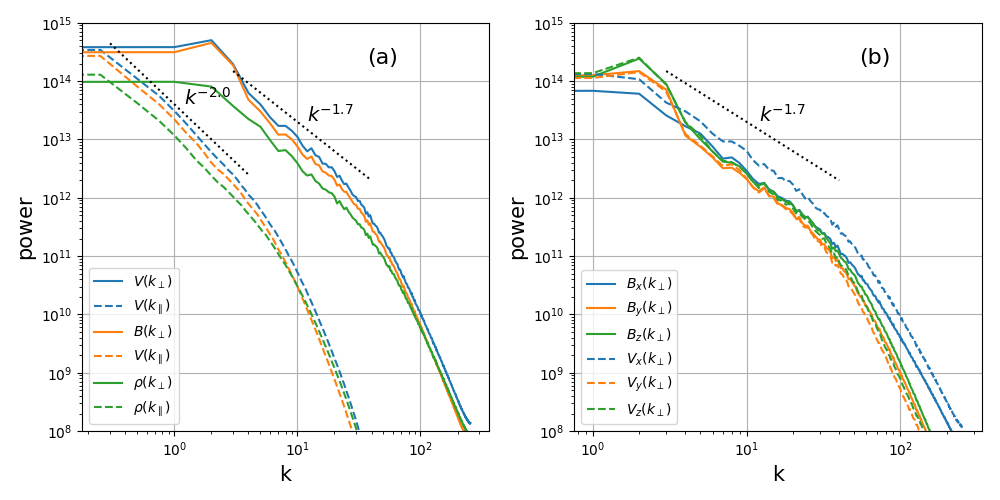}
    \caption{Power spectra of (a) total velocity field, magnetic field and density field fluctuations and (b) magnetic and velocity field components at $t=80$ for Run 1 . In the inertial range ($3<k<40$), both the magnetic field and velocity field power spectra follow approximately the Kolmogorov law with $k_\perp^{-5/3}$ in the perpendicular direction. 
    }
    \label{fig:spec_p}
\end{figure}

Figure \ref{fig:3dsnap} shows a surface contour of $\delta \rho$ at $t=80$ in Run 1. 
Clearly, $\delta \rho$ is anisotropic with much smoother variation along ${\bf B}_0$. 
Although the root-mean-square value of $\delta \rho$ is 0.13, there are localized 
structures with density enhancement or depletion with $|\rho-\rho_0|>0.4$.
The power spectra of $\delta \rho$ as a function 
of $k_\perp$ or $k_\parallel$ are shown in Figure \ref{fig:spec_p}(a). 
First, we observe that most of the power is contained in the perpendicular direction. 
It is same for the power spectra of magnetic field and velocity fluctuations, 
as shown in Figure \ref{fig:spec_p}(a), which is due to the anisotropic cascade of 
MHD turbulence in magnetized plasmas \citep{sheba_jpp_1983,GS95}. 
Second the perpendicular spectra have two power law segments with 
a break point near $k=100$, below which it follows a Kolmogorov law with $k^{-5/3}$, 
a signature of the inertial range.  The roll-off of the spectra above the break point 
is due to numerical dissipation at small scales. 
Third, fluctuations at $k \leq 3$ are mainly due to driving applied in the simulation.

In order to study how the density fluctuations will correlate with various
quantities, we decide to apply a bandpass to the density power spectra,
excluding fluctuations in both the injection range and the dissipation range. 
This is because the injection region dominates the spectral power, yet its
dynamics is subject to driving forces. 
We can calculate density fluctuations in the inertial range as, 
\begin{equation}
\overline{\delta \rho}\equiv \int_{k=k_{\rm inj}}^{k_{\rm ro}} \rho(k) dk~~.
\end{equation}
Similarly, we can calculate magnetic field and velocity field fluctuations 
in the inertial range as $\overline{\delta B}$ and $\overline{\delta v}$, respectively.
Note that the injection wave number $k_{\rm inj}$ is set by the driving forces ($k_{\rm inj}=3$ in all our simulations), 
and  the roll-over wave number $k_{\rm ro}$ is set by the numerical scheme for solving the MHD equations 
and the resolution of the simulation.
With other parameters being the same, Run 7 ($1024^3$ cells) 
has $k_{\rm ro}\sim 100$, Run 1 ($512^3$ cells) has  $k_{\rm ro}\sim 40$, and 
Run 8 ($256^3$ cells) has $k_{\rm ro}\sim 25$. But all three runs yield similar 
fluctuation levels in the inertial range, $\overline{\delta B}\sim 0.13$, 
$\overline{\delta v}\sim 0.14$ and $\overline{\delta \rho}\sim 0.08$. 
They are understandably smaller than the total fluctuations 
which include the injection range (cf. Table \ref{tab:para}).

Next, we study the relation of density fluctuation $\overline{\delta \rho}$ with 
turbulent Mach number $\overline{M_t}$ ($\equiv \overline{\delta v}/c_s$). 
Due to the limitation of computing resources, we use a large number of low-resolution 
($256^3$) simulations to cover the parameter space with $\beta=0.2, 0.5, 1.0, 2.0$, 
$\gamma=1.67, 2.70$, $\sigma_c=0.0, 0.3, 0.9$ and a range of $M_t$. 
The range of each parameter is inspired by observations. For example, 
PSP observations reported by \citet{nicol_apj_2020} suggest that the 
solar wind protons follow an equation state with a wide range of  $\gamma$, 
and the averaged $\gamma$ increases from $1.7$ at 1AU to $2.7$ in the inner heliosphere. 
High $\sigma_c$ and low $\beta$ are typical for the solar wind close to the Sun \citep[e.g.][]{kasper_prl_2021}, 
and $\sigma_c\sim 0$ and $\beta \sim 1$ are typical for the solar wind at 1 AU.

\begin{figure}
    \centering
    \includegraphics[width=0.48\textwidth]{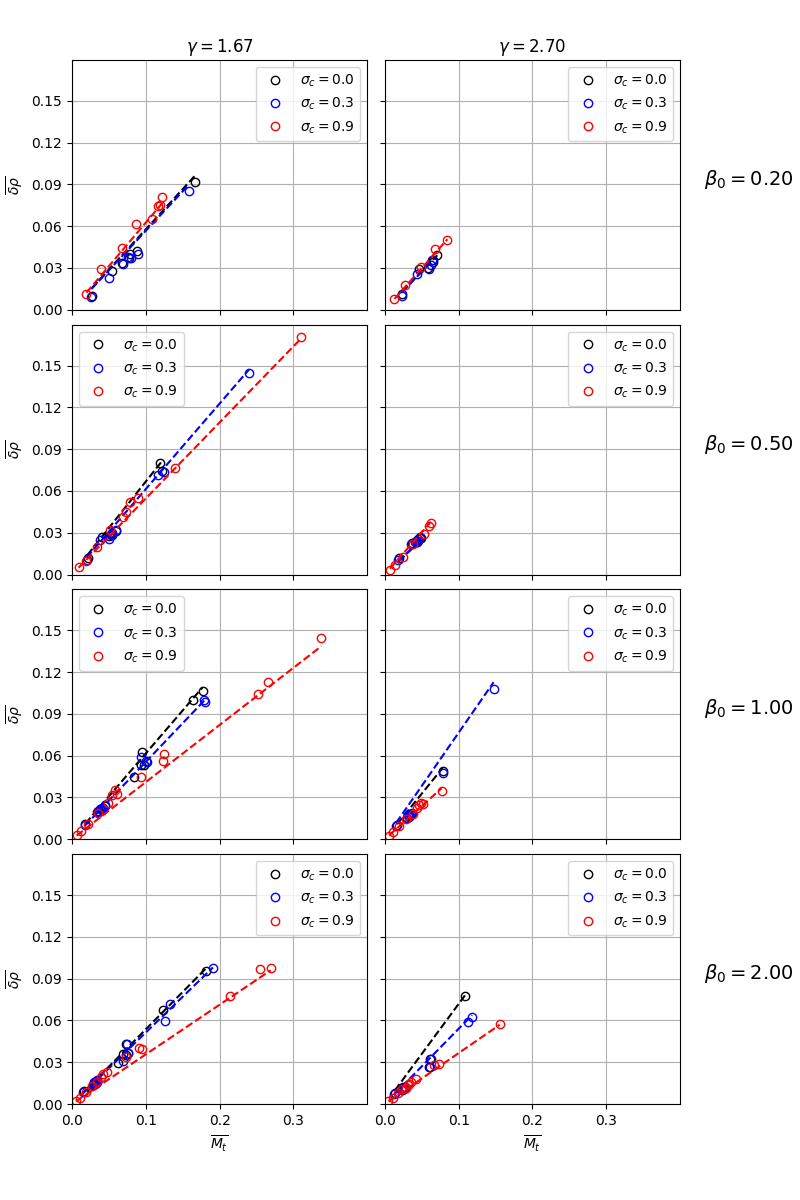}
    \includegraphics[width=0.48\textwidth]{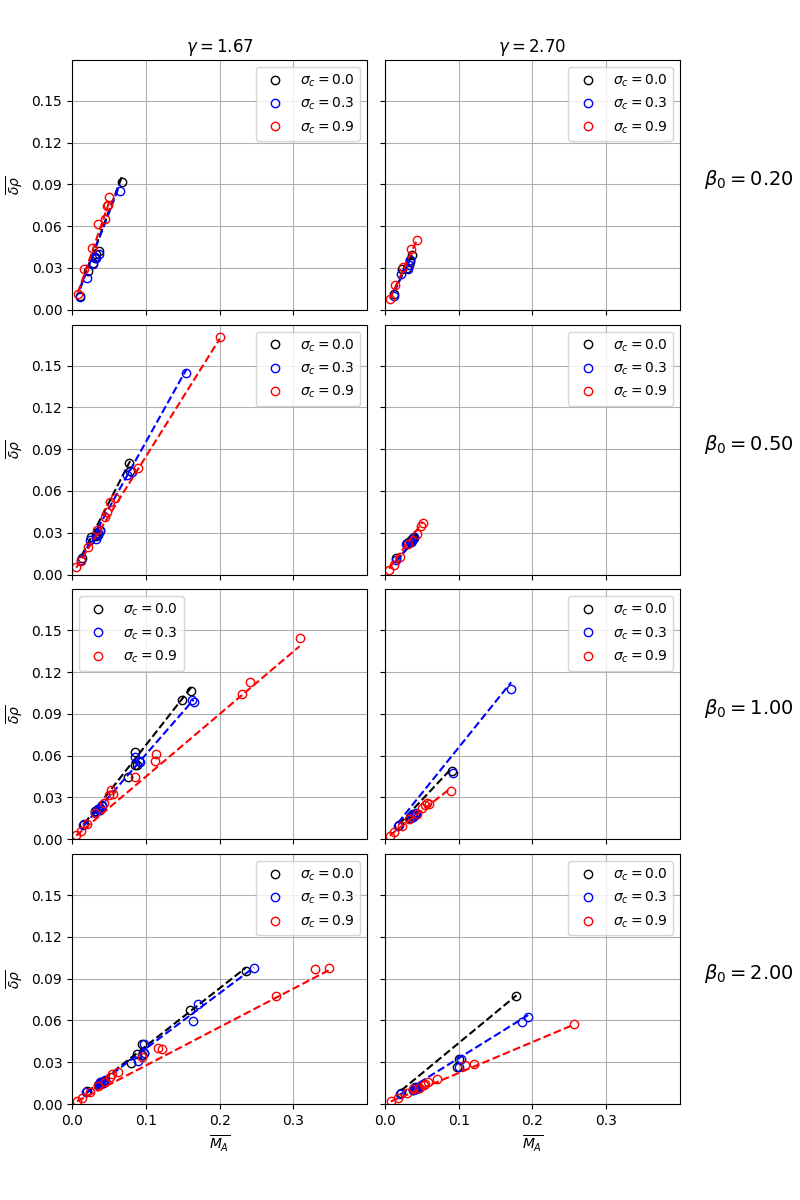}
    \caption{Density fluctuation $\overline{\delta \rho}$ as a function of (left) turbulent Mach number $\overline{M_t}$ and (right) Alfv\'en Mach number $\overline{M_A}$ for a series of 3D low-resolution ($256^3$ cells) MHD simulations with different combination of parameters $\beta$, $\sigma$ and $\gamma$. Circles are data points extracted from simulations, and dash lines are linear fit to these points. $\overline{\delta \rho}$ scales as  $C \overline{M_t}$, and the coefficient $C$ depends weakly on $\sigma_c$ and $\gamma$.}
    \label{fig:scaling2}
\end{figure}

Figure \ref{fig:scaling2} summarizes results from these low-resolution 
runs with different combinations of these key parameters. 
With other parameters ($\beta$, $\sigma_c$ and $\gamma$) fixed, 
we find $\overline{\delta \rho}$ scales nearly 
linearly as a function of $\overline{M_t}$, i.e.,
\begin{equation}
    \overline{\delta \rho} = C \overline{M_t}.
\end{equation}
The circles are data points extracted from simulations, and the dashed 
lines are linear fits to each set of runs. Note that although we 
have some control over $\sigma_c$,  it fluctuates in the simulations. 
So the numbers for $\sigma_c$ are not exact, but averaged values. 
$\beta_0$ is the initial value of $\beta$ at $t=0$, and $\beta$ 
increases slightly in simulation due to heating by injection, as discussed above.

Several features can be summarized from these results:
First, the coefficient $C$ has relatively weak dependence on $\beta$ when
using $\overline{M_t}$
as shown by comparing different rows in Figure \ref{fig:scaling2} (left panel).
If we plot the scaling of density as a function of Alfv\'en Mach number 
(right panel), there will be 
strong $\beta$ dependence, which is not surprising as 
$M_A\equiv M_t \sqrt{\gamma \beta/2}$. 
Second, the dependence of $C$ on both $\gamma$ and $\sigma_c$
is relatively weak. More balanced turbulence ($\sigma_c =0$) yields 
slightly higher density fluctuations than the imbalanced turbulence ($\sigma=0.9$). 
This trend is more prominent in runs with higher $\beta$ values. 
As we approach the Sun, generally $\beta$ decreases, $\gamma$ increases and 
$\sigma_c$ increases.  Changes of density fluctuation level can be 
complicated due to different effects of these parameters.  
Interestingly, for the same range of $M_t$, density fluctuations near the Sun might be
of similar magnitudes as those near 1 AU. 

\subsection{Nature of Density Fluctuations}

Lastly, we examine the nature of density fluctuations in our simulations. 
Traditionally, density fluctuations are associated with compressible waves.
Among the three linear MHD modes, Alfv\'en mode is incompressible, 
so only fast and slow modes can contribute to density fluctuations. 
Using linear mode decomposition in the spatial domain, 
fluctuations can be decomposed into the three MHD modes \citep{cho_prl_2002}. 
This decomposition is typically done on the velocity field, because the velocity 
fields from the three MHD modes form bases of a complete orthogonal system. 
We have performed such decomposition using frames near the end of Run 1.
We find that the mode decomposition method gives
about 75\% the velocity fluctuation power in the Alfv\'en mode, 
about 20\% in the slow mode and less than 5\% in the fast mode. 
The spatial linear mode decomposition approach, however,
does not take into account the temporal variations. For example, 
nonlinear structures, which do not belong to any linear mode, can still be decomposed 
spatially into three MHD modes. 
Therefore, as pointed out by \citet{gan_ApJ_2022}, this decomposition may 
overestimate the contribution from MHD waves. 

\begin{figure}
    \centering
    \includegraphics[width=0.9\textwidth]{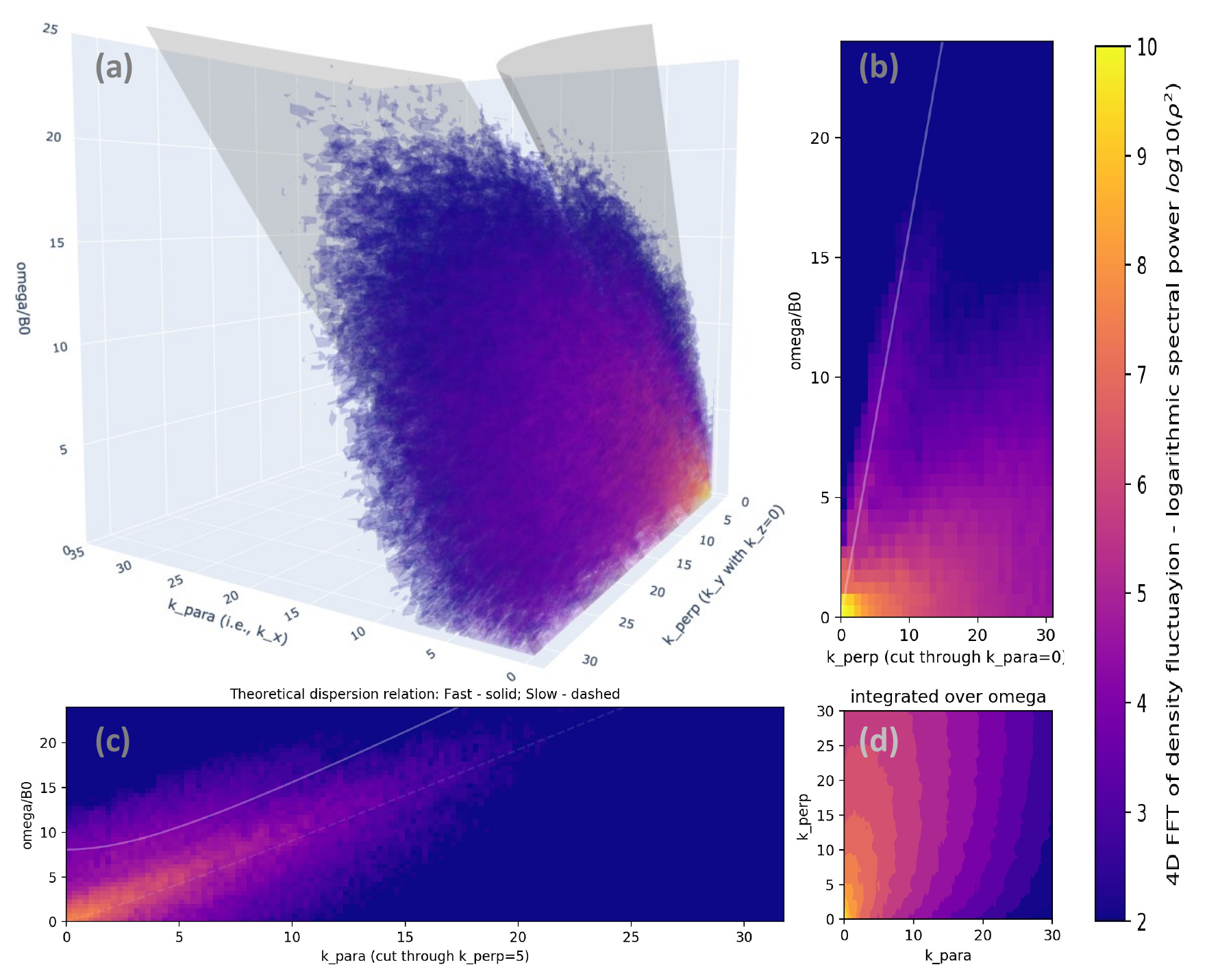}
    \caption{(a) Contours of the Fourier power of density fluctuations in the $\omega-{\bf k}$ space using 4D FFT analysis for Run 1, with $50<t<80$. For visualization purpose, the 4D FFT data has been reduced to 3D by projecting ${\bf k}$ vector into $k_\perp-k_\parallel$ plane. Grey surfaces represent the dispersion surface of the slow mode (lower) and the fast mode (upper). The majority of the power resides in oblique directions where $k_\perp\gg k_\parallel$ (c.f. Figure \ref{fig:spec_p}). The fractions of the power carried by the nonlinear structures, slow mode and fast mode are 91\%, 8\% and $<1$\%, respectively. (b) A 2D cut of the power in the $\omega-k_\perp$ plane with $k_\parallel=0$. The white line is the dispersion relation of the fast mode. (c) A 2D cut of the power in the $\omega-k_\parallel$ plane with $k_\perp=5$. The upper white line is the dispersion relation curve of the fast mode, and the lower white line is the dispersion relation curve of the slow mode. Most of the wave power is in the slow mode. (d) Fourier power of fluctuations in the $k_\perp-k_\parallel$ plane, showing strong anisotropy in the perpendicular direction. }
    \label{fig:4d_fft}
\end{figure}

To better analyze fluctuations in the simulations, we employ 
4D FFT analysis (3D for space and 1D for time), in which both 
spatial and temporal information are used to calculate the 
Fourier power in the $\omega-{\bf k}$ space \citep{gan_ApJ_2022}. 
Using this approach, the power along the dispersion surfaces of MHD waves 
 (numerically we sum over power within 10\% of theoretical frequencies) will be 
 regarded as ``waves", 
and fluctuations elsewhere will be called non-wave structures. 
Figure \ref{fig:4d_fft} shows such an analysis of the density fluctuations 
for Run 1, where we use data from the whole simulation domain 
in a time window $50<t<80$ (we exclude the initial growing phase). 
There is some power in two compressible modes: fast and slow modes. 
However, it is clear that the majority of the energy resides in the 
very low frequency structures, whose frequency is at or near zero and 
wave number is perpendicular to the background magnetic field 
(referred as ``low-frequency structure'' hereafter). This is 
more clear if we make a 2D cut in the $\omega-k_\perp$ plane with $k_\parallel=0$ (panel b). 
Summing up the power in the low-frequency structure, 
we find that it comprises nearly 91\% of the total power in the 
simulation box. MHD waves, on the other hand, contribute only to a 
small fraction, less than 9\%.
Furthermore, the majority of wave power resides in the slow 
mode (8\%) and the power of fast mode is negligible ($<1$\%), 
which can also be seen in Panel (c). Similar results were reported by 
\citet{gan_ApJ_2022}. Finally, Panel (d) shows the anisotropy 
of the density fluctuation, consistent with Figure \ref{fig:spec_p} (a). 

\begin{figure}
    \centering
    \includegraphics[width=0.9\textwidth]{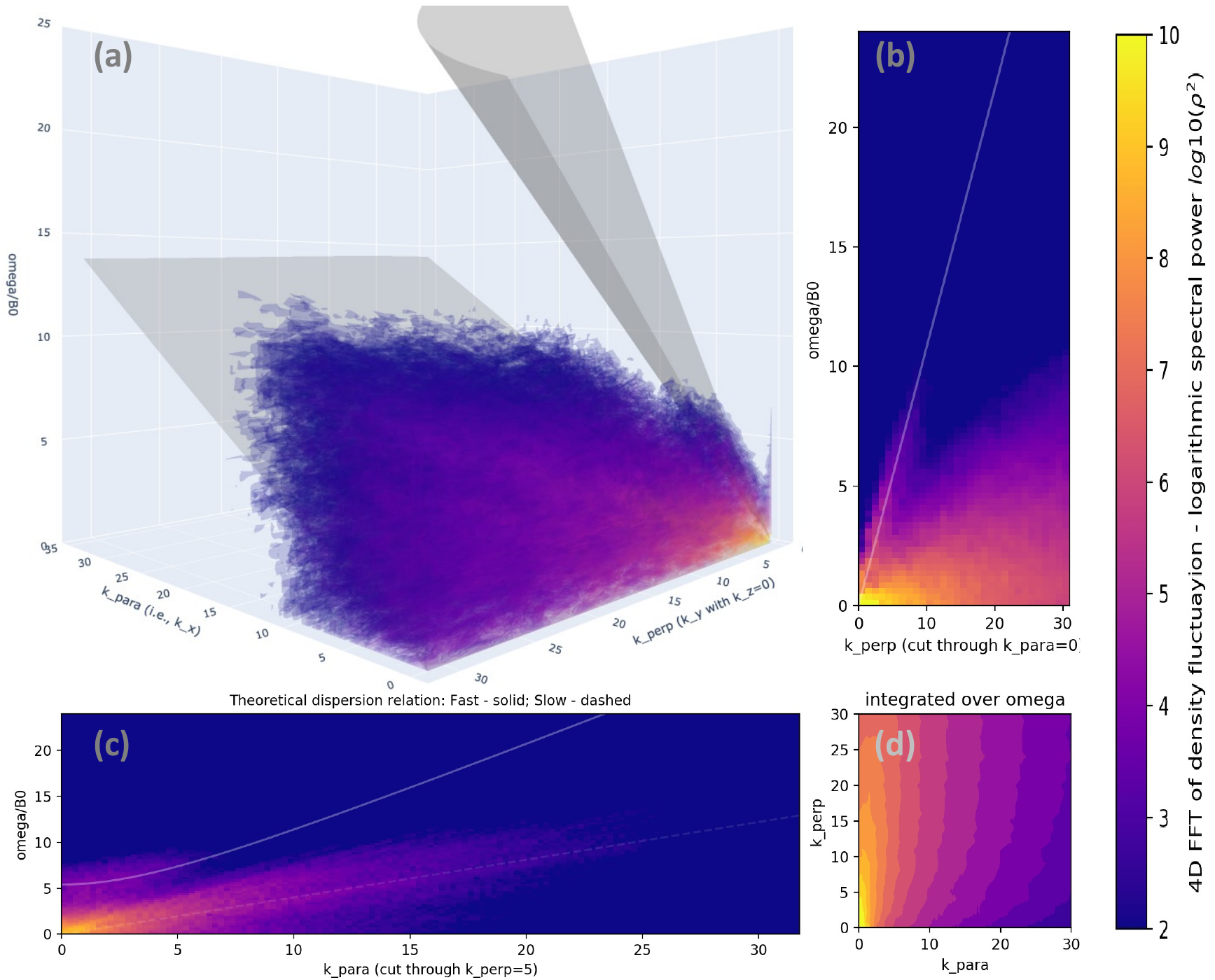}
    \caption{4D FFT analysis of density fluctuations in Run 4, with $t<60$. 97\% of the power is in the non-linear structures, and 3\% is in the slow mode. The contribution from the fast mode is negligible. The format is the same as Figure \ref{fig:4d_fft}. }
    \label{fig:4d_fft_0.2}
\end{figure}

Using 4D FFT analysis of the density fluctuations in the low-$\beta$ Run 4, 
we obtain qualitatively similar results. As in Figure \ref{fig:4d_fft_0.2}(a) and (c), 
the dispersion surface/curve of the slow mode 
($\omega=k_\parallel c_s=k_\parallel v_A \sqrt{\beta\gamma/2}$) is lower 
with a lower $\beta$. Still, the nonlinear structures contribute to 97\% of the 
total power, slow mode contributes 3\%, and the fast mode is negligible.

\section{Conclusion and Discussion} \label{sec:dis}
In this paper, using a large set of 3D MHD simulations, 
we investigate the development and properties of compressible turbulence 
driven by continuous large-scale perturbations. 
We have found:
\begin{enumerate}

    \item Although the driving (at large scales) is maintained incompressible throughout the simulations, compressible fluctuations of varying magnitudes 
    can still be generated via driving and they can be monitored via the RMS values of $\rho$, $\nabla \cdot {\bf v}$ 
    and $\nabla \cdot (\delta \rho {\bf v})$. In principle, compressible fluctuations can be 
    generated  via wave-wave interaction or mode conversion in an MHD system, 
    and it is likely that some of these processes are occurring (particularly for low $\beta$ simulations), 
    but we believe that the density fluctuations in our simulations are primarily due to driving, which is enforced 
    in both the momentum and magnetic fields. The turbulent Mach number in our simulations ranges from
    a few percent to the order of unity (including the driving range). After removing fluctuations in the driving range, 
    the turbulent Mach number drops down to be less than $0.3$.

     \item Simulations show a linear scaling of density fluctuation in the inertial range 
     as a function of turbulence Mach number $\overline{\delta \rho}= C \overline{M_t}$, 
     where the coefficient $C$ depends weakly on $\sigma_c$ 
     and $\gamma$. 
            
    \item Density fluctuations are dominantly caused by nonlinear structures with very low 
    frequency and large $k_\perp$. The nonlinear structures contribute to more than 90\%  
    of the fluctuation based on 4D FFT analysis whereas the actual compressible MHD waves
    contribute to a few percent in total power. 
    These nonlinear structures could be mistakenly attributed to MHD modes when using the 
    mode decomposition method based on the spatial variations 
    alone. The contribution from the fast modes is negligible.

\end{enumerate}

We have obtained a linear scaling of the density fluctuation 
as a function of turbulent Mach number based on our 3D MHD simulations. 
This is different from the prediction of nearly-incompressible MHD theory 
\citep{matth_pof_1988, zank_jgr_1992}, i.e., $\delta n\propto M_t^2$.  
The discrepancy may be caused by several reasons. First of all, NI-MHD theory 
assumes $M_t\ll 1$, whereas in our simulations $M_t$ ranges between $0.1 - 0.9$ (including the injection range).  
Second, NI-MHD theory attributes the density fluctuations to pseudo-sound 
which is a zero-frequency structure with $\nabla \cdot {\bf v}\approx 0$. As shown in our simulations, there are non-negligible $\nabla \cdot {\bf v}$ fluctuations and  the frequency of these 
nonlinear structures is low but not zero.
The scaling of density fluctuation was also compared with NI-MHD theory using solar 
wind data \citep{matth_jgr_1991, klein_jgr_1993, tu_jgr_1994, bavas_jgr_1995}, 
but the spreads in both $M_t$ and $\delta \rho$ data points were too large to draw
any firm conclusions. Note that, in-situ observation is 1D sampling through the solar wind turbulence, and the statistical properties are sensitive to sampling conditions such as the angle between the sampling line and the background magnetic field. They may also depend on plasma and turbulence parameters such as $\sigma_c$ and $\gamma$, as suggested by the simulations.  
Our linear scaling can be tested by observations if observational data 
can be grouped further by these parameters. Furthermore, 
how these results could vary depending on the driving processes needs
to be further investigated \citep{gan_ApJ_2022}. 

Finally, we have identified that the majority of the density fluctuations comes from 
nonlinear structures that do not follow the dispersion relation of linear MHD waves, 
based on 4D Fourier analysis. However, the exact nature of the structures and their generation 
mechanism are not fully understood. How these structures are 
related to the pressure-balance-structures 
\citep{vasqu_jgr_1999}, critically-balanced structures \citep{GS95} or psuedo-sound structures \citep{matth_jgr_1990} remains to be investigated.
We leave these topics to future investigation.

\acknowledgments
The research was supported by NASA under Award No. 80NSSC20K0377. The Los Alamos portion of this research was performed under the auspices of the U.S. Department of Energy. We are grateful for support from the LANL/LDRD program and DOE/OFES. Computing resources were provided by the Los Alamos National Laboratory Institutional Computing Program, which is supported by the U.S. Department of Energy National Nuclear Security Administration under Contract No. 89233218CNA000001.

%

\vspace{5mm}

\software{Athena++ \citep{stone_apjs_2020}, Matplotlib \citep{hunter_cse_2007}
          }





\bibliography{reference}{}
\bibliographystyle{aasjournal}


\end{CJK*}
\end{document}